\input{aipcheck}

\documentclass[
    ,final            
  ]
  {aipproc}

\layoutstyle{6x9}

\usepackage{graphicx,psfrag}
\usepackage{epsfig}
\usepackage{amsmath}
 \usepackage{amscd}
 \usepackage{amssymb}
\usepackage{epsf,float}

\def\beq{\begin{equation}}
\def\nn{\nonumber}
\def\eps{{\epsilon}}

\begin{document}
\title{Quantum Mechanics from Periodic Dynamics: the bosonic case}

\author{Donatello Dolce}{
  address={Johannes-Gutenberg Universit\"at,  D-55099  Mainz, Germany.}
}

 \begin{abstract}
	Enforcing  the periodicity hypothesis of the ``old'' formulation of Quantum Mechanics  we show 
	the possibility for a new scenario where Special Relativity  and Quantum Mechanics  are unified in a Deterministic Field Theory.  
		A novel interpretation of the AdS/CFT conjecture is discussed. 
	\end{abstract}
 
 \maketitle

	 \section{Periodic Dynamics}

	In the  ``old'' formulation of Quantum Mechanics (QM), free bosonic waves are supposed  to have  angular frequencies $\bar \omega  $; the energies of the related quanta $\bar E =  \hbar \bar \omega =h /T_t$ are fixed by  the inverse of the time periods $T_t = 2 \pi R_t$.
	We point out that to solve a relativistic differential system ($d s^{2}=c^2 d\tau^2 =  c^2 dt^2 - d\mathbf{x}^2$)  it is necessary to choose some Boundary Conditions (BCs). 
	The essential requirement is that the BCs  minimize the relativistic action  on the boundaries \cite{Csaki:2003dt}. For this reason in ordinary quantum field theory  one assumes fixed values of the fields on the time boundaries, let's say at $t$ and $t + T_t$. 
	For free bosonic fields, we  note that it is equivalently possible to satisfy the variational principle on the boundaries, and thus all the symmetries of the relativistic action, also by imposing Periodic BCs (PBCs) $\Phi(\mathbf x, t) \equiv \Phi(\mathbf x, t + 2 \pi R_{t}) $. This condition leads to remarkable overlaps with QM which we will briefly present in this  paper.  Details of the demonstrations,  further relevant issues and  physical interpretations are given in \cite{Dolce:2009ce}.
	
	Similarly to the harmonics of an acoustic field, by imposing PBCs as a constraint,  relativistic fields with different periodicities $T_t$ turn out to be decomposed as  towers of normal modes with discretized (quantized) angular frequency spectra $\omega_n = n \bar \omega = n/R_t$. 
	Assuming the de Broglie relation $\bar E = \hbar \bar \omega$  and in analogy with the Matsubara or the Kaluza-Klein (KK) theory, these fields can be naturally  interpreted  as  quantized towers of energy eigenmodes with energies proportional to the frequencies $E_n = n \hbar \bar \omega = n \hbar /R_t$. 
	 Supposing  on-shell fields, in the massless case ($d s^{2} = 0$) we must also consider an induced periodicity $\lambda_x =  2 \pi c R_t$ on the modulo of the spatial dimensions ($c^2 dt^2 = d\mathbf{x}^2$) and thus a quantization of the modulo of the momentum as well $|\mathbf p_{n}| = \hbar |\mathbf k_{n}| = n |\mathbf{ \bar p}|  =   {n \hbar }/{R_{t} c}$.  
	  Since the period is now related to the inverse of the fundamental energy $T_{t}(\mathbf{\bar p}) \equiv {h }/{\bar E (\mathbf{\bar p})} = {2 \pi }/{\bar \omega (\mathbf{\bar p})}$ through the Planck constant $h$, and not to the inverse of an invariant mass as in the KK theory, the periodicity must be regarded as dynamical. 
	 The space-time periodicities are described by the notation  $T_\mu = 2 \pi R_\mu = (R_t, \mathbf{R}_x/c) $ in order that the fundamental mode ($n=1$) has four-momentum $\bar p_\mu  
	 \equiv \hbar / c R^\mu$.  
	  We note that $T_\mu$ satisfies $\exp[-i x_\mu \bar p^\mu / \hbar ] \equiv \exp[-i (x_\mu +  c T_\mu) \bar p^\mu / \hbar ]$. 
	 For such a massless field  $\bar \omega (\mathbf{\bar p}) = {|\mathbf{\bar p}| c}/{\hbar}$, thus every  mode of the energy tower has  massless dispersion relation.
      	 In the limit of small fundamental momentum $|\mathbf{\bar p}| \rightarrow 0$ the periodicity tends to infinity $T_t \rightarrow \infty$, \textit{i.e.} the PBCs can be neglected and the continuos energy spectrum $\bar E \rightarrow 0$ is restored.	
	  Indeed  this is the usual relativistic field which  can also be obtained by putting $\hbar \rightarrow  0$. 
	 On the other hand,  in  analogy with the relativistic Doppler effect, increasing the momentum $|\mathbf{\bar p}| \rightarrow \infty$ through interaction or by changing the reference system, one obtains a relativistic deformation of all the space-time intervals and thus of the periodicities as well.     
	 This limit can be obtained by putting $\hbar \rightarrow \infty$, therefore it can be addressed as the quantum limit of the massless field. 
	  In fact in this case $R_t \rightarrow 0$, hence the PBCs are important and we get a well quantized energy spectrum $\bar E \rightarrow \infty$ at high frequencies. 
	  Actually, assuming that the  energy levels are populated according to the Boltzmann distribution (in a thermal field at temperature $ \mathrm T $ the $n^{th}$ level is populated proportionally to $\exp[- i E_n t /  K \mathrm T]$) we obtain a consistent description of the blackbody radiation, avoiding  the UV catastrophe \cite{Dolce:2009ce}. 
	 
	 Concerning  massive fields ($d s^2 > 0$), our main assumption is that it is possible to choose a reference frame where the proper time $\tau$ is equal to the real time: $ d\tau^2 =   dt^2$ (rest frame). 
	 This means that also the proper time has an  induced periodicity $ T_{\tau} \equiv   T_{t}(0) = h/ \bar M c^2$, where we define $\bar M$ to be the mass of the field.
	As can be shown by treating  the worldline parameter $s = c \tau$ as a \textit{virtual} eXtra-Dimension (XD)  for a 5-Dimensional (5D) field with zero 5D  mass, 
	that is $dS^2 =c^2 dt^2 - d\mathbf{x}^2 -  ds^2 \equiv 0$, we obtain again a regular energy tower $E_n(\mathbf {\bar p} ) = n \hbar \bar \omega(\mathbf {\bar p} )$; but now  every quantized level has the dispersion relation of a massive relativistic particle $\bar \omega(\mathbf {\bar p} ) =  {({\mathbf {\bar p}^{2} c^{2} + \bar M^{2} c^{4} })^{1/2}}/{\hbar}$, just as in the normal ordered second quantization,  thus in perfect agreement with Lorentz transformations \cite{Dolce:2009ce}. 
	Using our notation we find  $\bar M^2 c^4= h^2 / T^2_\tau  
	=(h /T^\mu)(h / T_\mu)$.
	The relation between these different de Broglie periodicities is actually $1/T^2_\tau = 1 / T^2_t - c^2 / \lambda^2_x$. 
	 In this way we see that the time periodicity $T_{t}(\mathbf{\bar p})$ has, in the rest  frame, an upper limit $  T_{\tau}$   proportional to the Compton wavelength $ \lambda_{s} \equiv T_{\tau} c = {h}/{\bar M c}$,  which indeed fixes the mass of the field: $E_n(0 ) = n \bar M c^2$. These fields can be thought of as \textit{dual} to XD fields with compactification lengths $\lambda_s$ (similarly to the KK tower, which has no tachyons, the quantized energy tower is positive defined  \cite{Dolce:2009ce}).
	 Even for a light boson at rest with the mass of an electron, this periodicity is already incredibly fast, $\sim 10^{-20} s$, if compared for instance with the characteristic period of the Cs-133 atom which by definition is of the order of $10^{-10} s$. 
	 Evidences of these intrinsic periodicities of the fields known as the ``de Broglie \textit{internal clocks}'' \cite{Broglie:1925}, have been indirectly observed only in a recent experiment for electrons \cite{2008FoPh...38..659C}. 
	 
	We want to point out that the theory  respects  relativistic causality, essentially because the PBCs minimize the relativistic action on the boundaries so that the periodicity $T_{t}(\mathbf{\bar p})$ is not statical but is as local and dynamical as the energy: $T_{t}(\mathbf{\bar p}) \equiv {h }/{\bar E (\mathbf{\bar p})} $.
	 Consequently, we can solve  the  Green function as usual for relativistic waves. 
	  Therefore, according to relativistic causality, by turning on a source in a point,  a retarded variation of the energy in a different point is induced and, just by energy conservation, a field in that interaction point passes from a periodic regime to another periodic regime. In this way it is possible to give a chronological order to events and the resulting  notion of time  formally 
	  fulfills  all the requirements of Special Relativity (SR) \cite{Dolce:2009ce}.  
	  
	 Since  we are in a  4D generalization of acoustic waves,  the whole information is contained in a single period. Indeed, in this non-trivial scenario we can use the terminology of field theory in XD and  say that time is compactified on a circle $t \in \mathbb{S}_{R_t}^{1}$ with radius $R_t$. 
	 
	 \section{Quantum Mechanics}

	We now want to study the mechanics of these periodic fields. For simplicity we assume only one spatial dimension. 
	The first thing we note is that the  field is a sum over on-shell standing waves, with Fourier coefficients $A_n$  ($n \in \mathbb{Z}$) which describe the population of the different energy levels, that is
	 \begin{eqnarray}\label{hilbert:H}
  \Phi(\mathbf x,t) =   \sum_{n} A_n  \phi_n( x) u_{n}(t)  = \sum_{n}  A_n  e^{-i (\omega_n t - k_n  x)} ~. 
  \end{eqnarray} 
Actually, we are in the typical case where a Hilbert space can be defined. 
	In fact,  the energy eigenmodes $\phi_n(x)$ form a complete set with respect to the  inner product $\left\langle \phi | \chi \right\rangle \equiv \int_{0}^{ \lambda_{x}} {d x}  \phi^* ( x) \chi( x) / {\lambda_{x}}$. 	 
	 Furthermore, the time evolution is described by the equations of motion $(\partial_t^2 + \omega_n^2)u_{n}(t) = 0 $, that can be reduced to first order differential equations  $i \hbar \partial_t \phi_{n}({x}, t) = E_n \phi_{n}({x}, t)$ \cite{Carena:2002me}.  
	 Formally, from the  Hilbert eigenstates $\left\langle {x}| \phi_n  \right\rangle_{} \equiv 
{ \phi_n({x}) }/{\sqrt{\lambda_{x}}}$  we can build the Hamiltonian operator: $\mathcal {H} \left| \phi_n \right\rangle_{}  \equiv \hbar \omega_n \left| \phi_n \right\rangle_{}$. 
	Similarly can be defined the momentum operator $\mathcal  P$. 
	{Thus, for an arbitrary state $| \phi  \rangle \equiv \sum_n a_n |\phi_n\rangle$  we have nothing else than the Schr\"odinger equation $i \hbar \partial_t |\phi\rangle =  \mathcal H |\phi\rangle$,} and the time evolution operator $ \mathcal U(t'; t) = \exp[{-{i}\mathcal {H}(t-t')}/{\hbar} ]$  is  Markovian:  $\mathcal U(t'';t') = \prod_{m=0}^{N-1} \mathcal U(t'+  t_{m+1}; t' + t_{m} -  \epsilon)$, where  $N \eps = t'' - t'~$ \cite{Nielsen:2006vc}. 
	 Without any further assumption than  periodicity we  plug the completeness relation in between the elementary time Markovian evolutions obtaining formally the Feynman Path Integral  (PI)   for  a time independent Hamiltonian \cite{Dolce:2009ce}
	  \begin{equation}\label{periodic:path.integr:Oper:Fey}
 U({x}'', t''; {x}', t') = \lim_{N \rightarrow \infty}   \int_{0}^{\lambda_{x}} \left ( \prod_{m=1}^{N-1} {d x_m}{ } \right )  \prod_{m=0}^{N-1}   \left\langle \phi  \right| e^{-\frac{i}{\hbar}( \mathcal {H} \Delta \eps_{m}  - \mathcal  P \Delta x_{m} ) }  \left| \phi \right\rangle_{},
\end{equation}
 	where  $\Delta x_{m} = x_{m+1} - x_{m}$ and $\Delta \eps_{m} = t_{m+1} - t_{m}$. 
	 In fact, also in the usual Feynman PI formulation the elementary evolutions are supposed to be on-shell \cite{Feynman:1948ur}
	\begin{eqnarray}\label{microevolut}
 & &U(x_{m+1}, t_{m+1}; x_m, t_m)  =\lambda_{x} \left\langle \phi  \right| e^{-\frac{i}{\hbar}( \mathcal {H} \Delta \eps_{m}  - \mathcal  P \Delta x_{m} ) }  \left| \phi \right\rangle_{} =   \nn \\ &=&  \sum_{n_{m}}  e^{- \frac{i}{\hbar} ( E_{n_{m}} \Delta \eps_{m} -  {p}_{n_{m}}  \Delta {x}_m)}  = {2 \pi}{} \sum_{n'}   \delta^{}\left( \bar \omega (\bar p) \Delta \eps_{m} -  \bar k \Delta x_{m}  + 2 \pi n' \right). 
\end{eqnarray}
	  	To write the last expression we have used the Poisson summation $\sum_{n}e^{-i n \alpha} = 2 \pi \sum_{n'}\delta(\alpha + 2 \pi n')$, therefore the PI eq.(\ref{periodic:path.integr:Oper:Fey}), or equivalently the periodic field itself eq.(\ref{hilbert:H}),  can be explicitly written as a sum over periodic on-shell paths 
		$U({x}'', t''; {x}', t' ) =  2 \pi   \sum_{n' } \delta^{}\left(\bar \omega (\bar p) (t'' - t') -   \bar k ({x''} - {x'})  + 2 \pi n'   \right )$. 
		Because the initial and final points are now defined modulo space-time periods they represent paths with different winding numbers. 
		 These on-shell Markovian and periodic paths can be ideally cut, translated by periods and combined  in such a way to form paths with the same initial and final space-time point \cite{Dolce:2009ce},  obtaining the  analogous of  the variations around a classical path of the usual Feynman formulation.  
	 In the original Feynman PI formulation there is  a single classical path linking the initial and final  point, so that  non classical paths must be necessarily considered to have interference.  
		 In our case the field can self-interfere because of the PBCs. In fact there is a set of periodic classical paths linking every initial and final configuration of the field, preserving the variational principle.   
	Nevertheless, as in the usual QM, we obtain that the  square modulo of the periodic field $|\Phi(x, t)|^2$ has a maximum along the relativistic free particle path where the periodic on-shell paths have constructive interference. 
    In the non-relativistic limit ${\bar p} \ll \bar M c$, the field can be thought of as at low intensity so that only the first level is largely populated \cite{Dolce:2009ce}. 
  Thus we  find  that the  periodic field itself eq.(\ref{hilbert:H}),  or equivalently the sum over periodic paths eq.(\ref{periodic:path.integr:Oper:Fey}), is reduced to the usual non-relativistic free particle distribution (modulo the de Broglie \textit{internal clock}) of the Feynman formulation:  $\phi( x, t) \sim A_0 \exp{[- i \frac{\bar M c^2}{\hbar}t + i \frac{\bar M}{ \hbar}\frac{x^2}{2 t}]} $. This also means that the field  is microscopically localized inside $\lambda_s$.  
   Hence \cite{Dolce:2009ce} we obtain a consistent interpretation of the wave/particle duality and of the double slit experiment. 
	 
	 The Heisenberg uncertainty relation can be easily obtained (we have standing waves) by absorbing the periodic invariance of the phase factor $\bar E t / \hbar + 2 \pi$  either as a time or as an energy variation: $\Delta E \times \Delta t = {(2 \pi \hbar)^{2}}/{ \bar E t} \geq  h$, where $t \leq T_t$. 
	 This is a direct consequence of the periodic conditions  $E_n {R_{t}} = n\hbar$ which can be stated in a  Bohr-Sommerfeld form: in a given potential the allowed solutions are those with integer numbers of cycles. 
	 Following this recipe it is easy \cite{Dolce:2009ce} to obtain the usual solutions of several non-relativistic Schr\"odinger problems.\footnote{The solution is given modulo a phase factors of ``little importance'' \cite{'tHooft:2001ar} in front of the wave functions which shift the energy spectrum (``still, no known phenomenon, including the Casimir effect, demonstrates that zero point energies are real'' \cite{Jaffe:2005vp}). It can be interpreted as a twist factor in the PBCs \cite{Dolce:2009ce}.}
	  Moreover,  generalizing  the symmetry breaking mechanism by BCs \cite{Csaki:2003dt} to a periodic electromagnetic  field at low temperature, we have an effective  magnetic flux quantization \cite{Dolce:2009ce} and other typical behaviors of  superconductivity \cite{Weinberg:1996kr}.  
	 
	 A further analogy with  canonical  QM emerges by noting that, from the given definition of  Hilbert space,  the expectation value of $\mathcal O(x)$ is 
	  \begin{eqnarray}\label{mean:value}
 \langle \chi(x_{f}, t_{f}) | \mathcal O(x) | \phi(x_{i},t_{i}) \rangle_{} \equiv
 \int_{0}^{\lambda_{x}} \frac{d^{} x}{\lambda_{x}} \sum_{n, m}  \chi^{*}_{m}(t_f,x_f) e^{  - i  k_{\chi m}  x}  \mathcal O(x)  e ^{ i  k_{\phi n}  x} \phi_{ n}(t_i,x_i)   ~.
\end{eqnarray} 
	 We suppose that the observable is $\mathcal O(x) \equiv \partial_x \mathcal F(x)$. Integrating by parts  taking into account the periodicity $\lambda_x$ of the spatial coordinate $x$,\footnote{ The theory can be equivalently reformulated by assuming Dirichlet BCs, \textit{i.e.}  $\Phi(0,t)=\Phi(\lambda_x,t)\equiv 0$ for spatial periodicity $\lambda_x$, eliminating the translational mode ($n=0$).} and  then imposing $\mathcal F(x)\equiv x$, we get $\langle \chi(x_{f}, t_{f})| 1 | \phi(x_{i},t_{i}) \rangle_{}   
 = \frac{i}{\hbar} \langle \chi(x_{f}, t_{f})| \mathcal  P  x -  x  \mathcal P    | \phi(x_{i},t_{i}) \rangle_{}$ \cite{Feynman:1948ur}. 
  For arbitrary initial and final periodic states $\phi(x_{i},t_{i})$ and $\chi(x_{f}, t_{f})$, this is  the usual commutation relation $[x,\mathcal P] = i \hbar$ of QM \cite{Dolce:2009ce}. 
 {We may also note that $[x, -i \hbar \partial_x]\Phi(x,t)= i \hbar \Phi(x,t)$. }  
	 
	 Indeed we have a  theory of relativistic waves where QM emerges because of periodic dynamics intrinsically too fast  with respect to our actual experimental time resolution. 
	 When the periodicity  is so fast, the system can only be  described statistically, since at every observation it   turns out to be in a random phase of its apparently aleatoric evolution. 
	 This is just like ``observing a clock under a stroboscopic light'' \cite{Elze:2002eg} or  a \textit{dice} rolling too fast to predict the result. 
	 In fact, we know from the 't Hooft determinism  that ``there is a close relationship between the quantum harmonic oscillator and the classical particle moving along a circle'' \cite{'tHooft:2001ar}. 
	In our case the are not local-hidden-variables being the time a physical variable that can not be integrate out, and being the PBCs an element of non locality. 
	 Therefore we can  speak about \textit{determinism} since the present theory is not constrained by the Bell's or similar theorems. 
	 
	 \section{AdS/CFT interpretation}
	 	 
	 To extend our theory to interacting fields we  could add an interaction terms in $\mathcal H$ of eq.(\ref{periodic:path.integr:Oper:Fey}), developing a perturbation theory. 
	  However, in simple cases, an approximative description of interactions can be intuitively formalized using a \textit{deterministic}  geometrodynamical approach.  
	 The trivial case is the  Compton scattering $e_i + \gamma_i \rightarrow e_f + \gamma_f$ where we  just rewrite the four-momentum conservation  among the quanta as the conservation of the reciprocal of the space-time periodicities  of the fields involved: $1/ T^{e_i}_{\mu} + 1/ T_{\mu}^{\gamma_i} = 1/  T^{e_f}_{\mu}+ 1/  T_{\mu}^{\gamma_f}$. 
	 We  see from this example that during interaction the periodicities are subject to deformations which in turn induce a corresponding deformation of the metric. Thus we formalize the problem of interaction using field theory in curved space-time. 
	 From the results obtained so far, we expect that the dynamics of these periodic fields in such a deformed background  reproduce the quantum behaviors  of the corresponding interaction scheme. Because of the underlying \textit{dualism} with an XD theory, this description can be regarded as a generalization of the AdS/CFT correspondence \cite{Maldacena:1997re}. 
	  For instance \cite{Dolce:2009ce}, we know from the Bjorken model  and the similitudes with a thermodynamic system \cite{Satz:2008kb} that the Quark-Gluon Plasma (QGP) can be  approximated  as  a volume of massless fields ($ds^{2} \simeq 0$) at high energy, and in particular that the energy decays exponentially \cite{Magas:2003yp} radiating hadronically and electromagnetically.  
	  Using our terminology and natural units, we get that the space-time periodicities $T_\mu$ of the fields have an exponential and conformal dilatation  $T_\mu  \rightarrow e^{ k s} T_\mu$ which can be alternatively described by the substitution $dx_\mu \rightarrow e^{- k  s} d x_\mu$. Indeed the QGP freeze-out is encoded in a AdS metric $dS^{2} \simeq  e^{- 2 k  s} dx_\mu dx^\mu - d  s^{2}$;  assuming $dS^{2} \equiv 0$ the proper time acts as a \textit{virtual} XD. 
	 As well known from AdS/QCD, the propagation of 5D YM fields with  5D coupling $g_5$ in  a  warped background gives a classical correlator $\Pi(q^{2})$ which, in first approximation\footnote{To evaluate the low energies effective propagator we use the holographic method with Neumann BCs at the UV scales $\Lambda$ and boundary field $A_{\mu}(q)$ at the IR scale $\mu$, in the hypothesis of Euclidean momentum $q$ such that $\Lambda \gg  |q| \gg \mu$. More intuitively we can use the ``zero mode approximation''. 
	 The leading contribution is given by the zero mode which in this case doesn't depend on $z= \exp[k s] / k$ so that $ \frac{A_0^{-2}}{4 g_{5}^{2}} \!  \int_{\Lambda^{-1}}^{\mu^{-1}} \!   \frac{d z}{k z} \!  \int \! d^{4} x F_{ \mu \nu}(x,z) F^{\mu \nu}(x,z) \! \!  \sim   \!\!  \frac{1}{4 g_{5}^{2}} \!  \int_{\Lambda^{-1}}^{\mu^{-1}}  \! \frac{d z}{k z}   \!  \int \! d^{4} x  F_{0 \mu \nu}(x) F_{0}^{\mu \nu}(x) \! \!  =  \! \! \frac{ \log \frac{\mu}{\Lambda}  }{4 k g_{5}^{2} }  \int \! d^{4} x F_{0 \mu \nu}(x) F_{0}^{\mu \nu}(x)$; hence the effective gauge coupling  $g$ turns out to have the logarithmic behavior of QCD $g^2 = {g_5^2 k}/{\log \frac{\mu}{\Lambda}} $ \cite{ArkaniHamed:2000ds}.} and  assuming ${1}/{k g_{5}^{2}}=  {N_{c}}/{ 12 \pi^{2}}$ \cite{Pomarol:2000hp,ArkaniHamed:2000ds}, can be matched  to the quantum vector-vector two point function of QCD:  $
  \Pi(q^{2}) \simeq   \frac{- q^{2}}{2 k g_{5}^{2}} \log \frac{q^{2}}{ \Lambda^{2}}  
  $. 
  In agreement with the interpretation \cite{Witten:1998qj} of AdS/CFT, from such an exponential and conformal dilatation of the space-time periodicities one actually obtains classically the quantum runnings of the strong coupling.

	\section{Conclusions}
	 
  Paraphrasing  the Newton's law of inertia and the de Broglie hypothesis \textit{we assume that  elementary free bosonic fields have intrinsic space-time periodicities $T_{\mu} \equiv { h }/{c \bar p^\mu }$.}
  These PBCs satisfy the variational principle  and  the theory is in agreement with SR. 
As much as the Newton's law  doesn't imply that every point particle goes in a straight line, our assumption does not  mean that the physical world should appear to be periodic. 
According to SR, these periodicities can vary through interactions (energy exchange) or by changing the reference system. Furthermore the combination of two or more periodic phenomena with irrational ratio of periodicities  results in \textit{ergodic} (nearly chaotic) evolutions.
  Remarkably, from this assumption of dynamical intrinsic periodicity the usual QM emerges under many of its  different formulations and for several nontrivial phenomena \cite{Dolce:2009ce}.  
	 This could open a new scenario where SR and  QM are \textit{unified} in a \textit{deterministic} field theory.
	 After all, the notion of time is strictly related with the assumption of periodicity: our usual -non compact- time axis is defined by counting the number of cycles of  phenomena \textit{supposed} to be periodic, in particular with reference  to  the Cs-133 atomic clock. 
	``We must assume, by the principle of sufficient reason'',\footnote{``\textit{By a clock we understand anything characterized by a phenomenon passing periodically through
identical phases so that we must assume, by the principle of sufficient reason, that all that
happens in a given period is identical with all that happens in an arbitrary period.}'' ~~~~~~~~~~~~~~~~~~~~~~~~~ A. Einstein \cite{Einstein:1910}}  periodicity to define a relativistic clock. 
Indeed, every elementary field can be regarded as having a relativistic de Broglie \textit{internal clock}. 
For massless (electromagnetic or gravitational) fields  these periodicities can in principle be infinite whereas in  massive fields they are bounded by the inverse of their masses. 
As in a calendar, the combination of the ``ticks'' of all these different \textit{internal clocks} is sufficient to fix \textit{uniquely} events in time and the usual external time axis can be dropped. 
 In these relativistic  \textit{internal clocks} ``all that happens in a given period is identical with all that happens in an arbitrary period''.$^4$ Thus, in a full relativistic generalization of acoustic fields,  every field can be regarded as characterized by dinamical compactified space-time dimensions. Massless fields with low frequency provide long space-time scales whereas non-relativistic massive fields  can be regarded as localized inside the Compton length, but with nearly infinite spatial period and microscopic time compactification, \textit{i.e.} as classical 3D  point-like particles.

\bibliographystyle{aipproc}

\end{document}